\documentclass[12pt]{article}
\begin{document}
\thispagestyle{empty}
\begin{center}
{\LARGE\textbf{Schr\"{o}dinger Equation for Nanoscience}}%

\vspace{1cm}
{\large Janina Marciak-Kozlowska\\
 Institute of Electron
Technology,
Al.~Lotnik\'{o}w~32/46 02--668~Warsaw Poland}%

\vspace{1cm} {\large Miroslaw Kozlowski\\
 Physics Department,
Science Teachers College  and Institute of Experimental Physics,\\
 Warsaw University,
Ho\.{z}a 69, 00-681 Warsaw Poland\\%
email: mirkoz@fuw.edu.pl}%
\end{center}
\vspace{2cm}
\begin{abstract}
The second order (in time) Schr\"{o}dinger equation is proposed.
The additional term (in comparison to Schr\"{o}dinger equation)
describes the interaction of particles with vacuum filled with
virtual particle~--~antiparticle pairs (\textit{zitterbewegung}).

\end{abstract}

\newpage
Quantum mechanics has been remarkably successful in all realms of
atoms molecular and solids. But even more remarkable is the fact
that quantum theory still continues to fascinate researches.
Interest in quantum mechanics both theoretical and experimental is
probably greater now that it ever has been.

 In this article we
develop the \textit{modified Schr\"{o}dinger} equation which
describes the structure of matter on the subatomic level i.e. for
characteristic dimension $r_n<d<r_a$ where $r_n\,(\textrm{nucleus
radius} )\sim\,\textrm{fm}$,
 $r_a\,(\textrm{atom radius})\sim~\textrm{nm}$. To
that aim we use the analogy between the Schr\"{o}dinger equation
and diffusion equation (Fourier equation). The quantum Fourier
equation which describes the heat (mass) diffusion on the atomic
level has the form~\cite{1}:
\begin{equation}
\frac{\partial T}{\partial
t}=\frac{\hbar}{m}\nabla^{2}T.\label{eq1}
\end{equation}
When the real time $t\rightarrow it/2$ and $T\rightarrow\Psi$,
Eq.~(\ref{eq1}) has the form of the free Schr\"{o}dinger equation:
\begin{equation}
i\hbar\frac{\partial \Psi}{\partial
t}=-\frac{\hbar^{2}}{2m}\nabla^{2}\Psi.\label{eq2}
\end{equation}
The complete Schr\"{o}dinger equation has the form:
\begin{equation}
i\hbar\frac{\partial \Psi}{\partial
t}=-\frac{\hbar^{2}}{2m}\nabla^2\Psi+V\Psi\label{eq3}
\end{equation}
where $V$ denotes the potential energy. When we go back to real
time $t\rightarrow-2it$, $\Psi\rightarrow T$ the new parabolic
quantum heat transport equation (quantum Fokker-Planck equation)
is obtained:
\begin{equation}
  \frac{\partial T}{\partial t}=\frac{\hbar}{m}\nabla^2
  T-\frac{2V}{\hbar}T.\label{eq4}
\end{equation}
Equation~(\ref{eq4}) describes the quantum heat transport for
$\triangle t>\tau$. For ultrashort time processes, $\triangle t
<\tau$ one obtains the generalized quantum hyperbolic heat
transport equation:
\begin{equation}
\tau\frac{\partial^2 T}{\partial t^2}+\frac{\partial T}{\partial
t}=\frac{\hbar}{m}\nabla^2 T-\frac{2V}{\hbar}T.\label{eq5}
\end{equation}
The structure and the solutions of Eq.~(\ref{eq5}) for ultrashort
thermal processes was investigated in the monograph: M.~Kozlowski,
J.~Marciak-Kozlowska: \textit{From quarks to bulk matter},
Hadronic Press, USA, 2001. The generalized heat transport
equation~(\ref{eq5}) leads to \textit{modified Schr\"{o}dinger}
equation~(MSE). After substitution $t\rightarrow it/2,
\,T\rightarrow\Psi$ in Eq.~(\ref{eq5}) one obtains:
\begin{equation}
  i\hbar\frac{\partial\Psi}{\partial
  t}=-\frac{\hbar^2}{2m}\nabla^2\Psi+V\Psi
  -2\tau\hbar\frac{\partial^2\Psi}{\partial t^2}.\label{eq6}
\end{equation}
The additional term (in comparison to Schr\"{o}dinger equation)
describes the interaction of electrons with surrounding space-time
filled with virtual positron-electron pairs, i.e.
\textit{zitterbewegung}.

One can conclude that for time period $\triangle t<\tau$ and
distance $\triangle r<c\tau$ the description of quantum phenomena
needs some revision. The numerical values for $\triangle t$ and
$\triangle r$ can be calculated as follows. Considering that
$\tau$,~\cite{1}
\begin{eqnarray}
  \tau&=&\frac{\hbar}{m\alpha^2 c^2}\sim 10^{-17}\,{\rm s}\\
  \label{eq7}
  c\tau&=&\frac{\hbar c}{m\alpha^2 c^2}\sim 1\,{\rm nm},\nonumber
\end{eqnarray}
we conclude that with the help of MSE we can visit the inner
atomic environement. On the other hand for $\triangle
t>10^{-17}$~s, $\triangle r>$~nm, in MSE the second derrivative
term can be omitted and as result the SE is obtained, i.e.
\begin{equation}
  i\hbar\frac{\partial \Psi}{\partial
  t}=-\frac{\hbar^2}{2m}\nabla^2\Psi+V\Psi.\label{eq8}
  \end{equation}
The visit of the inner structure of the atom can be quite
interesting, for the fact that atom radius remains strictly
constant during the universe expansion~\cite{2}.


\end{document}